# Topological layered *n*-type thermoelectrics based on bismuth telluride solid solutions

L. N. Lukyanova,[*] I. V. Makarenko, and O. A. Usov

Ioffe Institute, Russian Academy of Sciences, St Petersburg 194021, Russia

[*] E-mail: lidia.lukyanova@mail.ioffe.ru

In topological *n*-type thermoelectrics based on $Bi_2Te_3$ with atomic substitutions $Bi \to In$, $Te \to Se, S$, the morphology and the surface states of Dirac fermions on the interlayer (0001) surface of van der Waals were studied by scanning tunneling microscopy and spectroscopy (STM/STS) techniques. By the STM method, the dark and light spots on the surface were found, which intensities depend on the composition and thermoelectric properties of solid solutions. The observed surface morphology features in the solid solutions are explained by distortions of surface electronic states originated by atomic substitutions, influence of doping impurity, and formation of structural defects. Fast Fourier transform (FFT) of the morphology STM images of the (0001) surface were used to obtain the interference patterns of the quasiparticles excitation caused by surface electrons scattering by defects. The Dirac point energy and its fluctuations, peak energies of surface defects, the positions of the valence and conduction band edges, and the energy gap were determined from an analysis of tunneling spectra. A correlation between the parameters of surface states of Dirac fermions and thermoelectric properties was found. Thus, a contribution of the fermions surface states increases with rise of the surface concentration in solid solutions with high power factor, and the largest concentration value was observed in the $Bi_{1.98}In_{0.02}Te_{2.85}Se_{0.15}$ composition. The dependences of Fermi energy on the wave vector for different solid solutions are described by a set of Dirac cone sections, which are close within the limits of the fluctuations of the Dirac point energy that explained by weak changes of the Fermi velocity in the compositions at studied atomic substitutions in the bismuth telluride sublattices.

Keywords: bismuth telluride, solid solutions, topological insulator, Dirac point, Seebeck coefficient, thermoelectric power factor

## I. Introduction

Bismuth telluride based thermoelectrics are characterized by high figure of merit at operating temperatures of 100–500 K depending on the composition and charge carriers concentration and are widely used as thermoelectric energy converters for various purposes [1–3]. Topological phenomena promising for practical application in these thermoelectrics stimulate the studies of the anomalous properties of the topological surface states of Dirac fermions that possess characteristic linear dispersion and strong coupling between spin and momentum preventing fermions scattering by nonmagnetic impurities and defects [4–9]. The topological properties were most intensively studied for the binary compounds $Bi_2Te_3$ [10, 11], $Bi_2Se_3$ [12–14], and some solid solutions $Bi_2Te_2Se$ [15, 16], $BiSbTeSe_2$ [17]. The surface state parameters of the Dirac fermions have been determined from the analysis of transport properties in strong magnetic fields and by angle-resolved photoemission spectroscopy (ARPES) carried out for $Bi_2Te_3$ [18, 19], $Bi_2Se_3$ [20], and solid solutions $(Sb_{1-x}Bi_x)_2Te_3$ at x = 0.29 and 0.43 [21].

Currently, scanning tunneling microscopy (STM) and spectroscopy (STS) methods are used to study the surface electronic states of Dirac fermions [18, 22]. These methods allow to determine the features of surface morphology and the differential conductance proportional to the electron density of states, as well as Dirac point energy, the top of the valence band and the bottom of the conduction band, the position of the Fermi level, and the energy gap. Differential tunneling conductance was studied for binary compounds $Bi_2Te_3$ [18, 23], $Na_3Bi$ [24], $Bi_2Se_3$ [25], and for $BiSbTeSe_2$ [17], $Bi_{1.5}Sb_{0.5}Te_{1.7}Se_{1.3}$ [26] solid solutions. On STM images of the (0001) surface in [17, 23–26], locally charged *n*- and *p*-type regions caused by doping impurities were found. These charged regions can compensate the local fluctuations in the Coulomb potential, which is non-uniformly distributed on the (0001) surface [23, 24].

One of the specific features of topological thermoelectrics is the existence of some residual conduction in the bulk associated with defects. The optimal ratios between bulk and surface conductance can be achieved by defects compensation with variation of the thermoelectric material compositions [17, 27]. Investigations of the defects on the interlayer surface of $Bi_{2+x}Te_{2-x}Se$ solid solutions were carried out by STM method in Ref. [28] that allowed to reveal on morphology images the antisite defects of tellurium and bismuth due to excess of Te or Bi in the solid solution, respectively. Adsorbed impurities on the surface of the topological insulators of Bi and Sb chalcogenides and their influence on the topologically protected surface states of Dirac fermions were studied by STM/STS methods in $Bi_2Te_3$: (Ca, Mn) [23], $Bi_2Te_3$: (Ca, Rb, $NO_2$) [29], $Bi_2Te_3$: Rb, In, Ga, Au [30], and by the low-energy ion scattering (LEIS) method in $Bi_2Se_3$: Cs [31]. Analysis of the electronic structure of $Bi_2Te_3$ based on first principles calculations carried out with account of adsorption of Rb, In, Ga, and Au atoms on (0001) surface, show that the hybridization between the surface states of $Bi_2Te_3$ and nonmagnetic impurities can lead to the appearance of energy levels that depends on the chemical nature of the impurity [30].

The influence of topological surface states of fermions on the thermoelectric properties of topological materials has been considered in a number of modern aspects of scientific studies. Comprehensive analysis based on topology and thermal quantum field theory carried out in [9] show that topological surface charge carriers in $Bi_2Te_3$-based thermoelectrics lead to an anomalous increase in the Seebeck coefficient associated with local heat transfer along the temperature gradient as a result of the appearance of electron-hole Schwinger pairs on hot thermoelectric side. The dimensionless topological figure of merit *ZT* of 2.7 were obtained in Ref. [9] with account for only the electronic contribution, which is correlated with the experimental results for *p*-type $Bi_2Te_3/Sb_2Te_3$ superlattices [32].

In Refs. [6, 7], an increase in the figure of merit was related to the rise of the Seebeck coefficient caused by increase of the energy dependence of the spectral distribution of the mean free paths of both phonons and electrons in topological thermoelectrics originated from linear dispersion of fermions due to strong spin-orbit interaction. An enhance of the figure of merit due to an increase in mobility originated from the superfluidity effect of a topological exciton condensate that arises between *n*-type and *p*-type surfaces in heterostructures based on binary three-dimensional topological insulators $Bi_2Te_3$ and $Sb_2Te_3$ were considered in Refs. [33-36].

One of the effective methods for studying of the topological and thermoelectric properties of the materials is the external pressure method. Investigations of layered solid solutions of *n*-$Bi_2Te_{3-y-z}Se_yS_z$ and *p*-$Bi_xSb_{2-x}Te_3$ showed that the increase by 2–3 times of the power factor in the region of isostructural phase topological transitions at pressures of about 3 GPa is associated with rise of the mobility of charge carriers [37]. The observed enhance of the power factor at such pressures provides an increase in figure of merit at room temperature by about two times compared with normal conditions despite a rise in thermal conductivity of no more than 50% [38]. Similar thermoelectrics were used to fabricate *n*- and *p*-type legs in a model of thermoelectric module with adjustable mechanical stress [39, 40].

This work is devoted to STS/STM studies of topological *n*-type thermoelectrics (Bi, In)$_2$(Te, Se, S)$_3$ with atomic substitutions in the Bi and Te sublattices. An investigation of the (0001) interlayer surface by STM method with atomic resolution is used to analyze the influence of defects on surface morphology and interference pattern in the center of the Brillouin zone for various values of the Seebeck coefficient in the studied materials. The effect of the solid solution compositions on the contribution of the Dirac fermions surface states to thermoelectric properties is determined from the analysis of the differential tunneling conductance spectra *dI/dU* obtained by the STS method.

## II. EXPERIMENTAL PROCEDURES

Ingots of solid solutions $Bi_2(Te, Se)_3$, $Bi_2(Te, S)_3$, $(Bi, In)_2(Te, Se)_3$, and $Bi_2(Te, Se, S)_3$ were grown by directed crystallization method, which have been used to obtain multicomponent thermoelectrics of homogeneous composition. The studied samples were cut from single-crystal blocks of bulk ingots along the interlayer planes of van der Waals (0001) oriented along the growth axis perpendicular to the crystallographic axis of third order *c*. All samples crystallize in a rhombohedral crystal structure with five atoms in the primitive unit cell and the space group $R\bar{3}m\,(D_{3d}^5)$. The samples with a high Seebeck coefficient at $|S| > 270$ μV K$^{-1}$ have optimal thermoelectric properties for the low-temperature region of 100–220 K [41]. At higher temperatures above 300 K and up to 450–500 K, the optimal properties are observed at $|S| < 220$ μV K$^{-1}$. The power factor $S^2\sigma$ increases, while the Seebeck coefficient decreases accompanied with enhance of the electrical conductivity [42].

TABLE I. Seebeck coefficient S and thermoelectric power factor $S^2\sigma$ of *n*-type samples

| N | Composition | $S$ (μV K$^{-1}$) | $S^2\sigma^* 10^{-6}$ (Wcm$^{-1}$K$^{-2}$) |
|---|---|---|---|
| 1 | $Bi_2Te_3 + CdBr_2$ | –272 | 26.0 |
| 2 | $Bi_2Te_{2.7}Se_{0.3} + Te$ | –314 | 23.8 |
| 3 | $Bi_2Te_{2.7}Se_{0.15}S_{0.15} + Te$ | –272 | 31.4 |
| 4 | $Bi_2Te_{2.7}Se_{0.3} + CdCl_2$ | –185 | 44.2 |
| 5 | $Bi_2Te_{2.94}S_{0.06} + CdCl_2$ | –220 | 43.4 |
| 6 | $Bi_{1.98}In_{0.02}Te_{2.82}Se_{0.15} + CdCl_2$ | –213 | 46.3 |

*$\sigma$ - is the electrical conductivity.

Single-crystal samples of the solid solutions were investigated by STM/STS methods using a high-vacuum GPI-300 microscope (Prokhorov General Physics Institute RAS, Moscow) equipped with a high-vacuum module developed and manufactured by the Ioffe Institute, St.

Petersburg [43]. The interlayer surface (0001) of the studied sample with size of ~ $3\times4\times4$ mm$^3$ was fixed to the steel holder by electrically insulating epoxy adhesive. After that, the opposite interlayer surface of the sample was covered by adhesive tape to obtain clean surface by uplifting of the several layers from the corner of the crystal. The ohmic contacts both on the clean surface and on the holder were made by silver paste. Then, the sample was transferred through a system of gateways to a high-vacuum chamber, where using the Wobble Stick several upper layers of the sample were completely removed to obtain a fresh surface (0001). The measuring probes were made by etching of tungsten wire with a diameter of 260 $\mu m$ in 2M NaOH solution in order to reduce its curvature to 15 nm. Then in the preparation chamber the probes were annealed for several minutes at 600–700 °C at $10^{-6}$ Pa and subjected to Ar+ ion bombardment for 4 hours. STM/STS measurements were carried out in STM chamber at $2.4\times10^{-7}$ Pa at room temperature.

Morphology images of the (0001) interlayer surface with atomic resolution was measured by the STM method at a constant tunneling current mode of 0.1, 0.15, 0.2 nA (feedback turned on) and in voltage range (30–150 mV) in solid solutions, and up to 800 mV in Bi$_2$Te$_3$. The *dI/dU* spectra were measured by the STS method at preselected sample points in the constant height mode (feedback turned off) at a modulation voltage of 5–10 mV and a frequency of 7 kHz.

### III. RESULTS AND DISCUSSION

A. Morphology of the interlayer surface: Image analysis and Fast Fourier transform

The morphology images of the interlayer (0001) surface was measured by the STM for topological thermoelectrics of *n*-type Bi$_2$Te$_3$ and solid solutions with atomic substitutions Bi→In and Te→(Se, S) [Figs. 1(a)–1(f)].

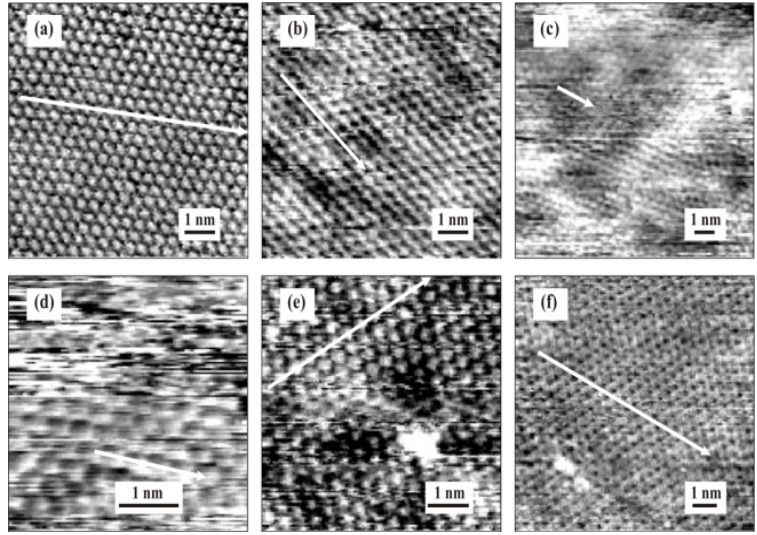

FIG. 1. The STM morphology images of the (0001) surface for *n*-type topological thermoelectrics based on bismuth telluride. The height differences (in Å) averaged over the surface for the studied samples are (a) 0.6 , (b) 0.5, (c) 0.7, (d) 0.2, (e) 0.4, (f) 0.4 for materials (a) $Bi_2Te_3$, (b) $Bi_2Te_{2.7}Se_{0.3}$ +Te, (c) $Bi_2Te_{2.7}Se_{0.15}S_{0.15}$, (d) $Bi_2Te_{2.7}Se_{0.3}$ +$CdCl_2$, (e) $Bi_2Te_{2.94}S_{0.06}$, (f) $Bi_{1.98}In_{0.02}Te_{2.85}Se_{0.15}$, respectively. The investigations of the profile height distribution along white arrows are shown in Fig. 2. The thermoelectric parameters are shown in Table I.

For all studied thermoelectrics, (0001) surface morphology is characterized by a hexagonal close-packed structure. The surface images in Fig. 1 show dark and light spots of various intensities observed for the most of the solid solutions compositions, which sizes are larger than the lattice constant and are about of 1–2 nm. Detected spots determine the waviness of the (0001) surface, which is associated both with distortions of surface electronic states at atomic substitutions in the Bi and Te sublattices, and formation of structural defects during the growth of solid solutions. The main defects in the considered thermoelectrics are antisite ones and vacancies. In samples containing excess Te [Table I(2), I(3)], the formation of Te antisite defects in the Bi sublattice is energetically most favorable. In the $Bi_2Te_{2.7}Se_{0.3}$ and $Bi_2Te_{2.7}Se_{0.15}S_{0.15}$ solid solutions, the spots related to defects on the (0001) surface were observed for samples with high Seebeck coefficients due to doping with excess Te [Figs. 1(b), 1(c)] [Table I]. In the Bi$_2$Te$_{2.7}$Se$_{0.3}$ sample with low Seebeck coefficient ($|S|$ = 185 μV K$^{-1}$), in which the defects are partly compensated by doping with the metal halide CdCl$_2$, the spots on the (0001) surface are less pronounced in comparison with the same composition of solid solution possessing high value of $S$ [Figs. 1(b)–1(d)]. Nevertheless, on the surface of the Bi$_2$Te$_3$ doped with the metal halide of CdBr$_2$ the spots were not observed despite of the high Seebeck coefficient [Fig. 1(a)] [Table I].

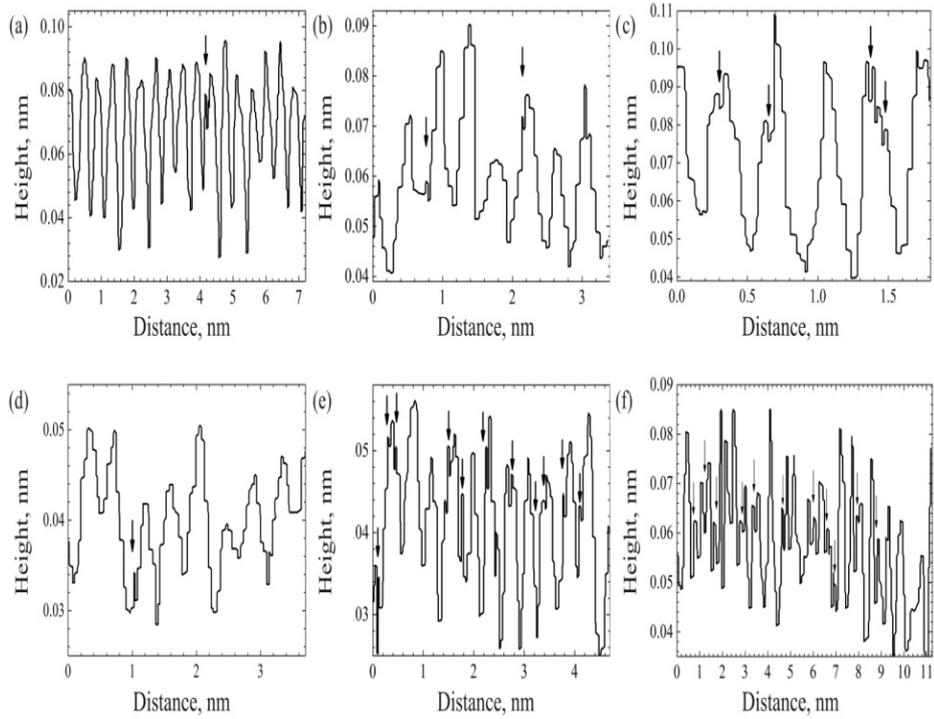

FIG. 2. The surface profiles describing height difference between neighboring atoms along the directions marked by white lines for the samples shown in Fig. 1. The observed splitting of peaks is marked by arrows.

In the Bi$_2$Te$_{2.94}$S$_{0.06}$ solid solution [Fig. 1(e)], even at small sulfur content, defects associated spots are formed on the (0001) surface, while spots in Bi$_2$Te$_{2.88}$Se$_{0.12}$ with low selenium content were not found [44]. An increase in the substituted atoms in the Te sublattice of the $Bi_{1.98}In_{0.02}Te_{2.85}Se_{0.15}$ multicomponent solid solution leads to the appearance of weak spots on the interlayer surface [Fig. 1(f)], as in the composition of $Bi_2Te_{2.79}Se_{0.21}$ [23, 44] with close total atomic substitution. Local bright spots observed on the (0001) surface of $Bi_2Te_{2.94}S_{0.06}$ + $CdCl_2$ and $Bi_{1.98}In_{0.02}Te_{2.85}Se_{0.15}$ + Te solid solutions in Fig. 1(e) and 1(f), respectively, can be related to adatoms formed both by dopants and substitutional sulfur or indium atoms.

The splitted peaks in Fig. 2 are related to atomic displacements at the sites of the crystal lattice, which increase in solid solutions compared with $Bi_2Te_3$ [Fig. 2(a)]. The peaks splitting increases in the $Bi_2Te_{2.7}Se_{0.15}S_{0.15}$ and $Bi_{1.98}In_{0.02}Te_{2.85}Se_{0.15}$ solid solutions with substitutions in both $Bi_2Te_3$ sublattices by S and In atoms possessing limited solubility [Figs. 2(c), 2(f)]. Despite a small number of sulfur atoms in the tellurium sublattice of $Bi_2Te_{2.94}S_{0.06}$, a large number of splitted peaks were also observed, which can be explained by the difference in the ionic radii of Te and S atoms [45].

The local values of the hexagonal in-plane lattice constant $a$, determined from profiles on the (0001) surface, are $\approx 4.06$ Å for $n$-$Bi_2Te_3$ [Fig. 2(a)] and $\approx 4.0$ Å for the $Bi_2Te_{2.94}S_{0.06}$ solid solution with a small number of substituted Te→S atoms [Fig. 2(e)]. The increase of substituted atom of Te→S and Bi→In leads to slight decrease of $a$ [Figs. 2(b)–2(d) and 2(f)], which is in qualitative agreement with those of X-ray diffraction data for $Bi_2Te_{3-y}Se_y$ solid solutions [46]. However, the values of $a$ in Fig. 2 are less than obtained by X-ray diffraction. According to [28, 47, 48] for $Bi_2Te_3$, $Bi_2Te_2Se$ and $Bi_2Se_3$ the values of the lattice constant $a$ in the hexagonal unit cell are 4.3835, 4.2975 and 4.1375 Å, respectively, and the composition dependence of $a$ is described by Vegard's law. Such discrepancies in the values of $a$ are caused by the differences of the experimental measurement methods because the local values of the lattice constant from the profiles [Fig. (2)] were determined with account of only atoms along the chosen direction on the (0001) surface.

As a result of the fast Fourier transform (FFT) of the morphology STM images of the (0001) surface, an interference patterns were obtained with the center at the point Γ of the Brillouin zone related to quasiparticle excitations during scattering of surface electrons by defects for some studied solid solutions that shown in Figs. 3(a)–3(d). The most pronounced interference pattern was observed for the $Bi_2Te_{2.7}Se_{0.3}$ + $CdCl_2$ solid solution [Fig. 3(a), insert], which corresponds to a high power factor [Table I] compared to other studied thermoelectrics.

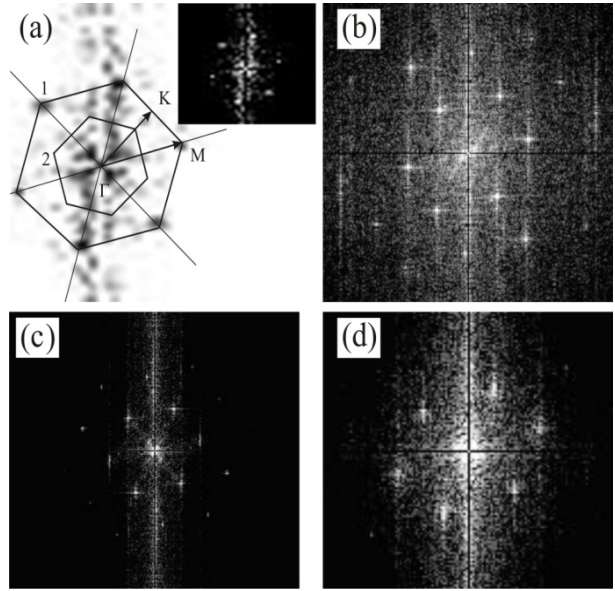

FIG. 3. Fourier transforms of (0001) surface images of topological thermoelectrics based on bismuth telluride. (a) $Bi_2Te_{2.7}Se_{0.3}$ + $CdCl_2$: the reciprocal lattice (1), the Brillouin zone (2) centered at the point Γ, and vectors ΓM and ΓK. The insert of (a): FFT image $Bi_2Te_{2.7}Se_{0.3}$ + $CdCl_2$, (b) $Bi_{1.98}In_{0.02}Te_{2.85}Se_{0.15}$, (c) $Bi_2Te_{2.94}S_{0.06}$, (d) $Bi_2Te_{2.7}Se_{0.15}S_{0.15}$.

From the analysis of the Fourier images of the (0001) surface, it follows that the reciprocal lattice vectors ΓM, ΓK, and the Brillouin zone constructed in Fig. 3(a) are similar for all other samples shown in Fig. 3. The first order spectral components are observed for the solid solutions $Bi_2Te_{2.7}Se_{0.3}$ + $CdCl_2$ [Fig. 3(a) insert] and $Bi_2Te_{2.7}Se_{0.15}S_{0.15}$ [Fig. 3(d)]. In addition to the first order spectral components the second and higher order ones are found for the solid solution $Bi_{1.98}In_{0.02}Te_{2.85}Se_{0.15}$ [Fig. 3(b)] and the second order ones for $Bi_2Te_{2.94}S_{0.06}$ [Fig. 3(c)]. It should be noted that the intensity of the high order components is larger for the compositions with atomic substitutions in both sublattices of bismuth telluride [Fig. 3(b)].

**B. SCANNING TUNNELING SPECTRA**

1. Fluctuations of the Dirac point energy

Dependencies of the differential tunneling conductance *dI/dU* on *U* were measured in the voltage range from –400 to 400 mV for the samples [Table I]. The energy of the Dirac point $E_D$ in *n*-type solid solutions corresponds to the minimum of the dependence of *dI/dU* on *U* [Fig. 4] and is located inside the energy gap for all studied samples in agreement with Ref. [17, 23, 27, 49].

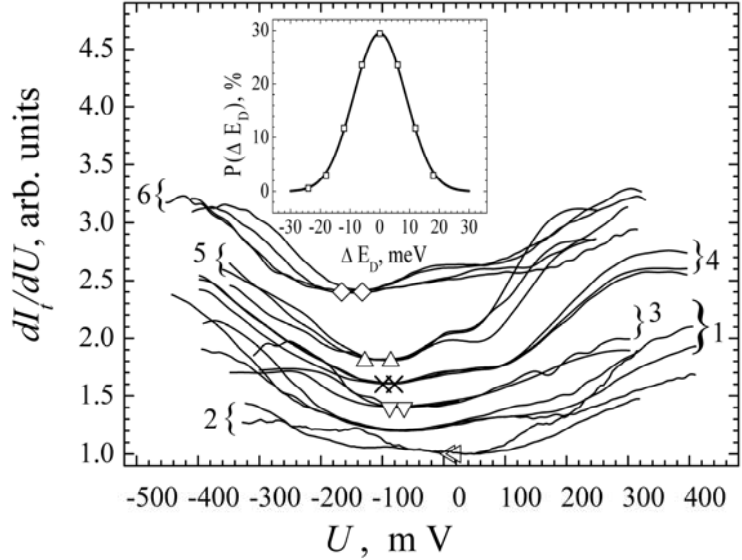

FIG. 4. Differential tunneling conductance *dI/dU* versus *U* measured at the (0001) surface for (1) $Bi_2Te_3$ + $CdBr_2$, and solid solutions (2) $Bi_2Te_{2.94}S_{0.06}$ + $CdCl_2$, (3) $Bi_2Te_{2.7}Se_{0.3}$ + Te, (4) $Bi_2Te_{2.7}Se_{0.15}S_{0.15}$ + Te, (5) $Bi_2Te_{2.7}Se_{0.3}$ + $CdCl_2$, (6) $Bi_{1.98}In_{0.02}Te_{2.85}Se_{0.15}$. All curves are normalized to the minimum value of conductance, and then the curves 2–6 are shifted up for clarity. The inset shows the normal Gaussian distribution function for Dirac energy fluctuations $P(\Delta E_D)$ versus $\Delta E_D = E_D - \langle E_D \rangle$, where $\langle E_D \rangle$ is the average energy of the Dirac point.

A series of *dI/dU* measurements versus *U* carried out on the same samples at different points of the (0001) surface showed the fluctuations $\Delta E_D$ of the Dirac point energy $E_D$ compared to its average value $<E_D>$ [Fig. 4, curves 1–6]. The energy fluctuations of the Dirac point are described by the normal Gaussian distribution [Fig. 4, inset], which is characterized by a standard deviation of about 8.9 meV for the studied samples, which is consistent with the results for $Bi_2Te_3$ and $Bi_2Se_3$ [23]. Fluctuations of $E_D$ in solid solutions is obtained to be significantly higher than for $Bi_2Te_3$ with $\Delta E_D$ of (–5.7) – (–4.8) meV [Fig. 4, curves 1]. A high level of fluctuations $\Delta E_D$ = (–22) – (18) meV was found in $Bi_2Te_{2.7}Se_{0.3}$ + $CdCl_2$ and $Bi_{1.98}In_{0.02}Te_{2.85}Se_{0.15}$ solid solutions, which exhibit high power factor [Figs. 4, curves (4), (6),] [Table I]. The fluctuations of $\Delta E_D$ are decreased in solid solutions of $Bi_2Te_{2.7}Se_{0.3}$+Te and $Bi_2Te_{2.7}Se_{0.15}S_{0.15}$ + Te with a high Seebeck coefficient [Figs. 4, curves (3), (5)] [Table I].

The features of $E_D$ fluctuations in solid solutions are correlated with images of the surface morphology (0001) and depend on the local spatial fluctuations of the Dirac point $E_D$ with respect to $<E_D>$. In $Bi_2Te_{2.94}S_{0.06}$ solid solution [Fig. 4, curve (2)] with intense spots observed on STM images [Fig. 1(e)], the ratio of the average fluctuation of the Dirac point $<\Delta E_D>$ to the average energy $<E_D>$ is about 32%. In the compositions of $Bi_2Te_{2.7}Se_{0.3}$+Te and $Bi_2Te_{2.7}Se_{0.15}S_{0.15}$ +Te [Figs. 4, curves (3), (4)] with high values of the Seebeck coefficient, the $<E_D>$ is varied from -77 to -92 meV, respectively, the ratios $<\Delta E_D>/<E_D>$ is decreased to 10%, and the spots intensity in the images of surface morphology weakens [Figs. 1(b), 1(c)]. The observed fluctuations of the Dirac energy $\Delta E_D$ at various points on the (0001) surface were determined by the inhomogeneous distribution of the Coulomb potential [17, 23, 24, 50] caused by distortion of the surface atomic structure related to a local change in the composition of the solid solution both due to substitutions of Bi→In, Te→Se atoms and the influence of excess Te and metal halide doping.

## 2. Energy peak positions in STS spectra

Besides, to the Dirac point energy fluctuations, the peaks $p_i$ corresponding to surface levels formed by defects were found in Fig. 5 and Fig. 6. The peak energy $E_P(E_D)$ can be written as $E_P(E_D) = E_P - E_D$, since the energy of surface defect levels measured with respect to the Dirac point $E_D$ does not depend on the position of the Fermi level as was shown in Ref. [21, 34]. For the solid solutions of $Bi_2Te_{2.7}Se_{0.3}$ + Te, $Bi_2Te_{2.7}Se_{0.15}S_{0.15}$ [Figs. 5, curves (2), (3)] the values of $E_P$ are negative and for the $Bi_2Te_{2.94}S_{0.06}$ + $CdCl_2$ [Fig. 5, curve (1)] the relative peak energies are found both as negative ($E_{p1}$, $E_{p2}$) and positive ($E_{p3}$, $E_{p4}$).

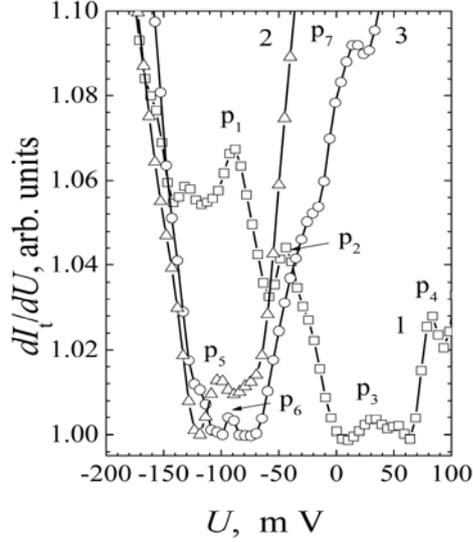

FIG. 5. Normalized differential conductance ($dI/dU$) versus voltage $U$ in solid solutions (1) $Bi_2Te_{2.94}S_{0.06}$, (2) $Bi_2Te_{2.7}Se_{0.3}$ + Te, and (3) $Bi_2Te_{2.7}Se_{0.15}S_{0.15}$. The peaks $p_i$ corresponding to the energy of defects $E_{p_i}$ with respect to the Dirac point for curves 1–3 are: (1) $Ep_1$ = -105 meV, $Ep_2$ = -58 meV, $Ep_3$ = 15 meV, $Ep_4$ = 70 meV, (2) $Ep_5$ = -25 meV, and (3) $Ep_6$ = -5 meV, $Ep_7$ = 106 meV.

As follows from Figs. 5 and Fig. 6, the intensity and position of the peaks are influenced by the composition of the solid solution, Seebeck coefficient, and power factor values. In the $Bi_2Te_{2.94}S_{0.06}$ + $CdCl_2$ solid solution with a low $S$-value and a high power factor, the peak intensity and energy are higher than in $Bi_2Te_{2.7}Se_{0.15}S_{0.15}$ + Te with a high $S$-value and a low power factor [Table I].

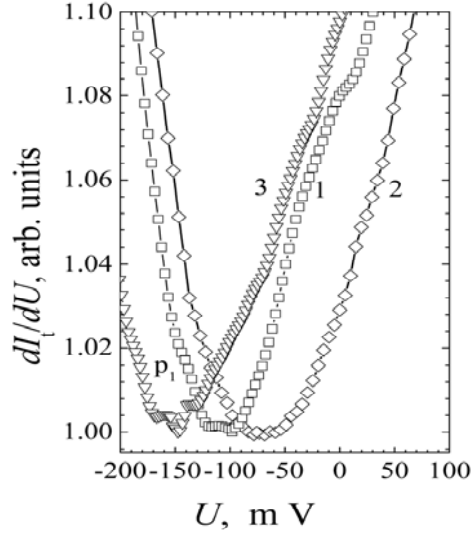

FIG. 6. Normalized differential conductance ($dI/dU$) versus voltage $U$ for (*1*) $Bi_2Te_3$, solid solutions (*2*) $Bi_2Te_{2.7}Se_{0.3}$ + $CdCl_2$, and (*3*) $Bi_{1.98}In_{0.02}Te_{2.85}Se_{0.15}$. The peak energy obtained for $Bi_{1.98}In_{0.02}Te_{2.85}Se_{0.15}$ is $Ep_1$ = -8 meV with respect to the Dirac point.

Besides, the observed decrease in the intensity and energy of the peaks is accompanied by reduce of spots contrast on (0001) surface morphology images [Fig. 1(c)]. In the $Bi_2Te_3$ and $Bi_2Te_{2.7}Se_{0.3}$ + $CdCl_2$ compositions [Figs. 6, curves (1), (2)], defect associated peaks on the differential conductance curves were not detected, while the spots in the images were weak for $Bi_2Te_{2.7}Se_{0.3}$ + $CdCl_2$ [Figs. 1(d)] and were not pronounced at all for $Bi_2Te_3$ [Fig. 1(a)].

3. Surface concentration of Dirac fermions

From the differential tunneling spectra plotted in Fig. 5 and Fig. 6, the energies of the top of the valence band $E_V$ and the bottom of the conduction band $E_C$ were determined from the positions of the inflection points on the normalized tunneling curve $dln(I)/dln(U)$ measured in a wider voltage range $U$ from –400 to 400 mV in accordance with Ref. [44, 49]. The obtained energies of the Dirac point $E_D$ (curve 1), the edges of the valence band $E_V$ (curve 2) and the conduction band $E_C$ (curve 3) with respect to the Fermi energy $E_F$ = 0 (curve 4) and the energy gap $E_g$ (curve 5) depending on the Dirac fermions surface concentration $n_s$ are presented in Fig. 7. The $n_s$ is determined in accordance with the expressions [51]

$$n_s = \frac{1}{4\pi} k_F^2 \qquad (1)$$

where the wave vector $k_F$ is equal to 

$$k_F = \frac{|E_D|}{v_F} \qquad (2)$$

and $v_F$ is the Fermi velocity.

The Fermi velocity was determined in accordance with the Vegard's law using data for the B$_2$Te$_{3-y}$Se$_y$ solid solutions system at y = 0 [52], 0.9 **[53]**, 1 [54, 55], 3 **[56]**, since $v_F(y)$ is close to linear dependence in the range of studied compositions. A similar dependence of the Fermi velocity on the composition of solid solutions (B$_{1-x}$Sb$_x$)$_2$Te$_3$ was also observed in Ref. [57].

For Bi$_2$Te$_{2.7}$Se$_{0.15}$S$_{0.15}$ composition, calculations of wave vector $k_F$ were carried out using $v_F$ for of Bi$_2$Te$_{2.7}$Se$_{0.3}$, since the total atomic substitution in the Te sublattice in these thermoelectrics was the same. However, ARPES studies [58, 59] of the Bi$_{1.1}$Sb$_{0.9}$Te$_2$S$_1$ show that large Te→S atomic substitutions leads to the increase of the Fermi velocity. Evidently, a small amount of atomic substitutions Te→S in the Bi$_2$Te$_{2.7}$Se$_{0.15}$S$_{0.15}$ solid solution is weakly affect the $v_F$ value. The $E_F(k_F)$ dependences in Bi$_{1.98}$In$_{0.02}$Te$_{2.85}$Se$_{0.15}$ and Bi$_2$Te$_{2.94}$S$_{0.06}$ solid solutions with small additions of indium and sulfur atoms were also determined with account for Fermi velocity $v_F$ of Bi$_2$Te$_{3-y}$Se$_y$ solid solution system.

Since the Dirac point in the investigated *n*-type thermoelectrics is located in the energy gap, the increase of energy $|E_D|$ leads to the shift of the Dirac point towards the top of the valence band that is related to an enhance of the surface concentration $n_s$ in dependence on the solid solution composition [Fig. 7, curve (1), points 6–11].

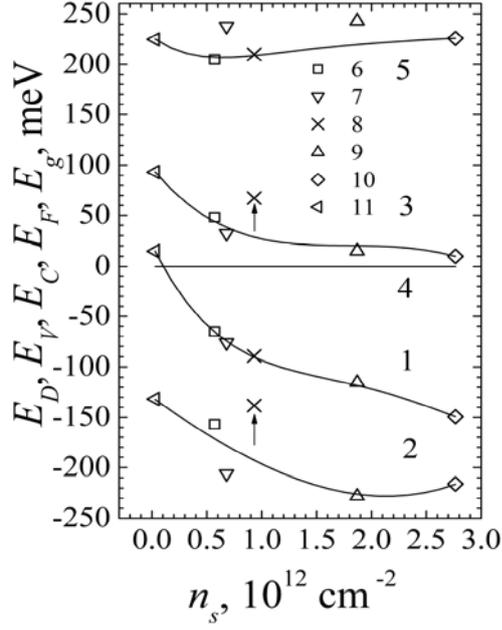

FIG. 7. The average energy of the Dirac point $E_D$ (1), the position of the edges of the valence band $E_V$ (2), the conduction band $E_C$ (3) with respect to the Fermi energy $E_F = 0$ (4) and the energy gap $E_g$ (5) versus fermions surface concentration $n_s$. The points corresponds to the experimental energy parameters for $Bi_2Te_3$ (6) and solid solutions $Bi_2Te_{2.7}Se_{0.3}$+Te (7), $Bi_2Te_{2.7}Se_{0.15}S_{0.15}$ (8), $Bi_2Te_{2.7}Se_{0.3}$+$CdCl_2$ (9), $Bi_{1.98}In_{0.02}Te_{2.85}Se_{0.15}$ (10), $Bi_2Te_{2.94}S_{0.06}$ (11). The arrows indicate the deviations of the $E_V$ and $E_C$ values from curves 2, 3.

In $Bi_2Te_{2.7}Se_{0.3}$+Te and $Bi_2Te_{2.7}Se_{0.15}S_{0.15}$ solid solutions with high Seebeck coefficients and low volume concentration of charge carriers in bulk samples of $\approx 3 \times 10^{18}$ cm$^{-3}$ [42] the surface concentration of fermions $n_s$ grows weaker than in $Bi_2Te_{2.7}Se_{0.3}$+$CdCl_2$ and $Bi_{1.98}In_{0.02}Te_{2.85}Se_{0.15}$ with high power factors [Table I] and high bulk concentration of $\approx 2 \times 10^{19}$ cm$^{-3}$ [41], [Fig. 7, points 7–10]. The highest increase of the $n_s$ and the largest shift of the Dirac point energy to the top of the valence band $E_V$ were observed in the $Bi_{1.98}In_{0.02}Te_{2.85}Se_{0.15}$ solid solution [Fig. 7, point 10] with $n_s = 2.8 \times 10^{12}$ cm$^{-2}$ of one order of magnitude higher than the composition of $Bi_2Te_{2.94}S_{0.06}$ [Fig. 7, point 11], which also possess a high power factor [Table I]. Thus, in addition to a high power factor, an increase of the surface concentration $n_s$ depends on the composition of the solid solution. It should be noted that in the $Bi_2Te_{2.94}S_{0.06}$ solid solution, the $E_D$ is found above the Fermi level $E_F$ that results in the hole surface conductance of the Dirac fermions of *p*-type [Fig. 7, point 11], while for all other studied samples the Dirac point $E_D$ [Fig. 7, points 6–10] is located below Fermi level [Fig. 7, curves (1), (4)].

The shift of the $E_V$ and $E_C$ bands edges in the studied materials leads to an increase in the energy gap $E_g$ compared to the optical data, which is explained by variation of the electron density of states at inversion of the energy gap edges of topological insulators [60]. The value of $E_g$ increases slightly in $Bi_2Te_{3-y}Se_y$ solid solutions with increase of substituted atoms in the Te sublattice. However, in the composition of $Bi_2Te_{2.94}S_{0.06}$, the value of $E_g$ becomes close to $Bi_2Te_{2.7}Se_{0.3}$ one despite the small number of substituted atoms in the Te sublattice [Fig. 7, curve (5)].

4. Dirac cone sections

Investigations of the STS spectra in thermoelectrics of various compositions allow to determine the Fermi energy $E_F$ with respect to the Dirac point depending on the wave vector $k_F$ calculated according to equations Eq. [1, 2].

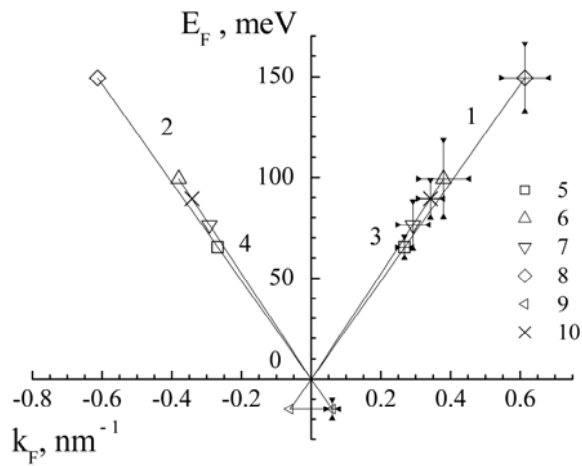

FIG. 8. Dependence of the Fermi energy $E_F$ on the wave vector $k_F$ (lines 1–4) for (5) $Bi_2Te_3$ and solid solutions (6) $Bi_2Te_{2.7}Se_{0.3}$ + $CdCl_2$, (7) $Bi_2Te_{2.7}Se_{0.3}$ + Te, (8) $Bi_{1.98}In_{0.02}Te_{2.85}Se_{0.15}$, (9) $Bi_2Te_{2.94}S_{0.06}$, and (10) $Bi_2Te_{2.7}Se_{0.15}S_{0.15}$. The dependences of $E_F(k_F)$ demonstrate the ranges of $E_F$ variation associated with local fluctuations in the energy of the Dirac point $\Delta E_D$ [Fig. 4] and the corresponding changes of $k_F$.

The coincidence observed in Fig. 8 between curves (1), (3) and (2), (4) of the linear dependences of $E_F(k_F)$ for the studied thermoelectrics in the fluctuation range of $E_F$ is related with a small change in the Fermi velocity for the investigated atomic substitutions in the Te sublattice. Thus, the dependences of the Fermi energy $E_F$ on the wave vector $k_F$ for the thermoelectrics of various compositions [Fig. 8] are a set of Dirac cone sections superimposed on each other due to a weak change of the Fermi velocity.

## IV. Conclusion

The morphology of the van der Waals interlayer surface (0001) and the differential tunneling conductance $dI/dU$ versus voltage $U$ were studied by STM and STS in the $n$-type solid solutions based on $Bi_2Te_3$ with atomic substitutions of Bi→In, Te→Se, Se + S, S. The morphology of the (0001) interlayer surface, which contains dark and light spots of various intensities depending on the composition and thermoelectric properties of solid solutions, was studied by STM method. Increase of the spots intensity was observed in the thermoelectrics which possess a high Seebeck coefficient and, in particular, with atomic substitutions of Bi and Te by limited solubility components, such as indium and sulfur. From the height profiles of the (0001) surface obtained along the chosen directions, the splitted peaks, which characterize the features of atomic displacements at crystal lattice sites, were found. It was observed that the splitting of peaks, as much as spots intensities, increases in solid solutions with components of limited solubility. An analysis of STM images of (0001) surface morphology of solid solutions by the FFT method were allowed to reveal the interference patterns of the quasiparticles excitation that originate from the surface electrons scattering by defects. The most clear quasiparticle interference pattern was obtained for the $Bi_2Te_{2.7}Se_{0.3}$ + $CdCl_2$ solid solution with a high power factor. The revealed morphological features of the (0001) interlayer surface of solid solutions are explained by probable distortions of electronic surface states that originate from atomic substitutions as a result of doping impurities influence and structural defects formation.

By STS method, the energy of the Dirac point $E_D$, the positions of the valence $E_V$ and conduction $E_C$ band edges, the Fermi level $E_F$, and the energy gap $E_g$ in thermoelectrics of various compositions were determined. It was shown that the intensity of local fluctuations in the Dirac point energy $\Delta E_D$ and surface levels of defects associated peak energies near the minimum $dI/dU$ were found to be correlated with the Seebeck coefficient and power factor in the solid solutions.

The enhance of the Dirac fermions surface states contribution to thermoelectric properties were revealed in the solid solutions $Bi_{1.98}In_{0.02}Te_{2.85}Se_{0.15}$ and $Bi_2Te_{2.7}Se_{0.3}$ doped with $CdCl_2$ possessing high values of the power factor, in which the surface concentration of Dirac fermions $n_s$ is higher than in other studied thermoelectrics. The maximum contribution of surface states was observed in the $Bi_{1.98}In_{0.02}Te_{2.85}Se_{0.15}$ solid solution with the value of surface concentration $n_s$ up to $2.8 \times 10^{12}$ cm$^{-2}$, which is one order of magnitude higher than for $Bi_2Te_{2.94}S_{0.06}$.

It was found that the dependences of the Fermi energy $E_F$ on the wave vector $k_F$ for investigated solid solutions with account for influence of fluctuations of the Dirac point energy $\Delta E_D$ are described by a combination of Dirac cone sections superimposed on each other due to insignificant changes in the Fermi velocity $v_F$ for atomic substitutions in the sublattices of bismuth telluride.

**Acknowledgments**

This study was partially supported by Russian Foundation for Basic Research Project No. 20-08-00464.